\documentclass[12pt]{iopart}

\usepackage{epsfig}
\usepackage{graphicx}

\newcommand{\kb}{k_{\mathrm{B}}}
\newcommand{\nub}{\nu_{\mathrm{B}}}
\newcommand{\rhom}{\rho_{\mathrm{m}}}

\newcommand{\ud}{\/\mathrm{d}\/}
\newcommand{\ple}{P}
\newcommand{\pss}{P^{(\mathrm{ss})}}
\newcommand{\JH}{J_\mathrm{H}}
\newcommand{\Pone}{P^{(1)}}
\newcommand{\Ptwo}{P^{(2)}}
\newcommand{\Pthree}{P^{(3)}}
\newcommand{\Qtwo}{Q^{(2)}}
\newcommand{\qtwo}{q^{(2)}}
\newcommand{\PN}{P_N}

\begin{document}
\title[Derivation of Fourier's law in stochastic energy
exchange systems] 
{On the derivation of Fourier's law in stochastic energy
  exchange systems} 
\author{Pierre Gaspard\textdagger, Thomas Gilbert\textdaggerdbl}
\address{Center for Nonlinear Phenomena and Complex Systems,\\
  Universit\'e Libre  de Bruxelles, C.~P.~231, Campus Plaine, B-1050
  Brussels, Belgium}

\begin{abstract}
We present a detailed derivation of Fourier's law in a class of
stochastic energy exchange systems that naturally characterize 
two-dimensional mechanical systems of locally confined particles in
interaction. The stochastic systems  
consist of an array of energy variables which can be partially exchanged
among nearest neighbours at variable rates. We provide two independent
derivations of the thermal conductivity and prove this quantity is
identical to the frequency of energy exchanges. The first derivation
relies on the diffusion of the Helfand moment, which is determined solely
by static averages. The second approach relies on a gradient expansion 
of the probability measure around a non-equilibrium stationary state. The
linear part of the heat current is determined by local thermal equilibrium
distributions which solve a Boltzmann-like equation. A numerical scheme is
presented with computations of the conductivity along our two methods. The
results are in excellent agreement with our theory.
\end{abstract}

\submitto{J. Stat. Mech.}


\ead{\textdagger gaspard@ulb.ac.be, \textdaggerdbl thomas.gilbert@ulb.ac.be}

\section{\label{sec.intro}Introduction}

A key to a comprehensive derivation of transport properties starting from a
microscopic theory is to identify the conditions under which scales in a
macroscopic system. This problem is especially 
challenging for interacting particle systems. To establish the conditions
under which scales separate is to provide an understanding as to how a
large number of particles whose microscopic motion is described by Newton's
equations organise themselves in irreversible flow patterns at the
macroscopic level.

In the context of thermodynamics, macroscopic equations such as Fourier's
heat law derive from conservation laws supplemented by phenomenological 
ones. The former concern say the conservation of mass or energy, and reflect 
the existence of the corresponding
conservation laws at the level of Newton's equations. Phenomenological laws
on the other hand provide linear relations between the currents
associated with the flow of conserved quantities and
thermodynamic forces in the form of gradients of the conserved quantities,
thereby introducing a set of transport coefficients. Though
the values of these coefficients can be precisely measured and are usually
tabulated for the sake of their use in the framework of applied
thermodynamics, they cannot be otherwise determined  
without knowledge of the underlying dynamics. A first-principles based
computation of the transport coefficients consequently requires a deep
knowledge and understanding of the dynamics, as well as
their statistical properties.  

For this purpose, a common procedure in non-equilibrium statistical
physics --see \emph{e.g.} Uhlenbeck's discussion of Bogoliubov's approach
to the description of a gas in \cite{Kac} or van Kampen's views on the
role of stochastic processes in physics \cite{VK}--
is to apply the two-step programme which consists of 
(i) identifying an intermediate level of description --a mesoscopic scale--
where the Newtonian dynamics can be consistently approximated by a set of
stochastic equations, and (ii) subsequently analyzing the statistical 
properties of this stochastic system so as to compute its transport
properties.

The first part of this programme was successfully completed in a recent set
of papers \cite{GG08a,GG08b}, in which we introduced a class of Hamiltonian
dynamical systems describing the two-dimensional motion of locally confined
hard-disc particles undergoing elastic collisions with each other. Whereas
the confining mechanism prevents any mass transport in these systems, energy
exchanges can take place through binary collisions among particles
belonging to neighbouring cells. Under 
the assumption that binary collisions are rare compared to wall 
collisions, it was argued that, as a consequence of the rapid decay of
statistical correlations of the confining dynamics, the global
multi-particle probability distribution of the system  typically 
reaches local equilibrium distributions at the kinetic energy of each
individual particle before energy exchanges proceed. This mechanism
naturally yields a stochastic description of the time evolution of the
probability distribution of the local energies in terms of a master
equation to be described below. An important property in that respect is
that the accuracy of this stochastic reduction can be controlled to
arbitrary precision by simply tuning the system's parameters. 

In \cite{GG08a,GG08b}, arguments were presented supporting the result that
the heat conductivity associated with this master equation has a simple
analytical form, given by the frequency of energy exchanges. In a sense,
this result is similar to the fact that, for instance, uniform random walks
describing tracer dynamics have diffusion coefficients given in terms of
jump probabilities of the walkers. However, 
contrary to such systems, the stochastic system at hand lacks a special
property which facilitates the derivation of such results, namely the
gradient condition. A system obeys the gradient condition if there
exists a local function such that the current, whether of mass or energy,
can be written as the difference of this function evaluated at separate
coordinates \cite{Spo91}. In such cases, the diffusion coefficient is
determined through a static average only, and is therefore easy to
compute. Given that our stochastic system does not verify the gradient
condition, it is in fact remarkable that the heat
conductivity should take such a simple form.

In this paper, we complete the second part of the programme described above
and provide a systematic derivation of the heat conductivity associated
with heat transport in the class of stochastic systems derived in
\cite{GG08a,GG08b}. We justify in particular why, in spite of the fact that
our systems do not obey the gradient condition,  the heat conductivity can
still be computed from the Green-Kubo formula through a static average
only. We achieve this by an alternative method which consists in
considering the stationary heat flux produced by a temperature gradient
across the system. We suppose the system is a two-dimensional slab. Along
the first dimension, the system has finite extension, with the
corresponding borders in contact with stochastic thermal baths at different
temperatures. The system may be taken to periodic along the second
dimension. As a result, a temperature gradient develops along the first
dimension, which induces a heat flux from 
the warmer border to the colder one. We then set up a scheme to compute the
resulting non-equilibrium stationary state and  obtain the linear relation
connecting the temperature gradient to the heat flux. This scheme proceeds
by consistently solving the 
Bogoliubov-Born-Green-Kirkwood-Yvon (BBGKY) hierarchy up to pair
distributions through a type of Chapman-Enskog gradient expansion. As the
heat current associated with our systems depends on neighbouring pairs
only, the knowledge of the pair distribution function to first order in the
temperature gradient makes possible the computation of the heat current in
the linear regime and, consequently, that of the thermal conductivity,
thereby completing the derivation of Fourier's law. 

Our theoretical results are furthermore supported by the results of
direct numerical simulations of the stochastic system, with excellent
agreement. Simulation methods of kinetic processes governed by a master 
equation have been extensively developed after Gillespie's original work 
\cite{Gil76,Gil77}. Here we describe a method which accounts for continuous 
energy exchanges among all the pairs of neighbouring cells.

The paper is organised as follows. The master equation is described in
section \ref{sec.meq}, with relevant definitions. Statistical objects are
described in section \ref{sec.ss}, identifying equilibrium and
non-equilibrium stationary states. Local temperatures are defined in
section \ref{sec.en} and the energy conservation law derived, thereby
identifying heat currents. Section \ref{sec.helf} offers a first
computation of the heat conductivity through the method of Helfand
moments. An alternative computation is presented in section \ref{sec.ce},
solving the BBGKY hierarchy as sketched above. The numerical scheme for the
computation of the heat conductivity is discussed in section \ref{sec.nr},
with a presentation of the results. Finally, conclusions are drawn in
section \ref{sec.con}.

\section{\label{sec.meq}Master equation}

\begin{figure}[htb]
  \centering
   \includegraphics[width = .75\textwidth]{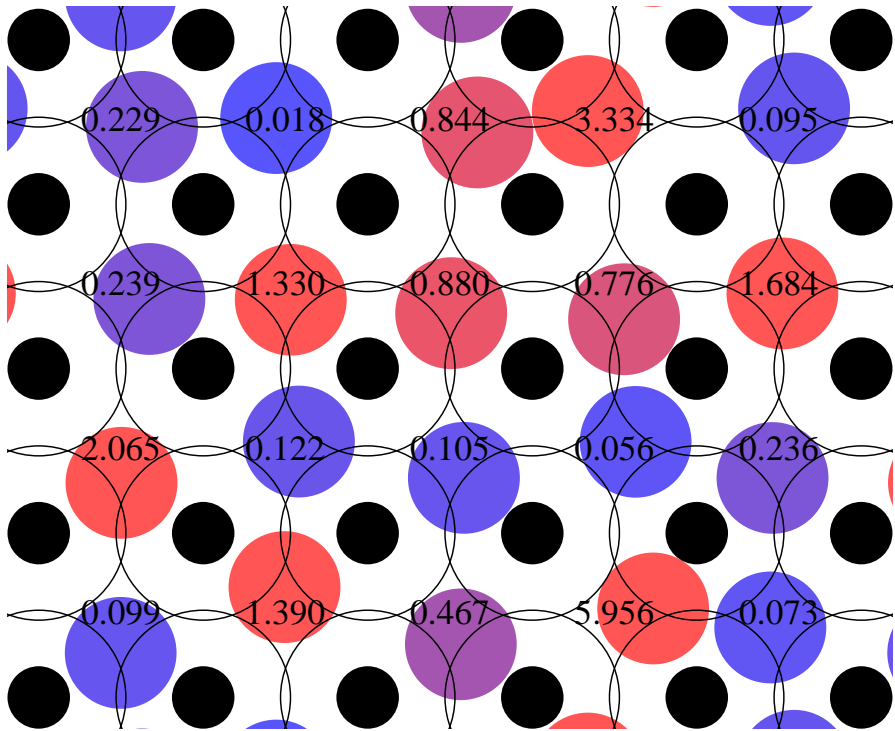}
   \caption{Snapshot showing an example of an interacting particle system with a confining
     mechanism, such as described in \cite{GG08b}. The coloured particles,
     colour-coded according to their kinetic energies from blue to red,
     move among an array of fixed black discs. The system has two 
     parameters, the diameters of the fixed and moving discs. They are chosen
     so that their sum is larger than the distance between neighbouring
     fixed discs. This guarantees that the moving discs are trapped within
     their cells; their
     centers move within semi-dispersing billiards of bounded horizon
     (exterior intersection of the black circles). Though they are
     confined, the moving discs can nevertheless exchange energy through
     binary collisions as long as their diameters are properly chosen. The
     numbers indicate the values of the kinetic energies in every cell. In
     the limit where binary collisions are rare with respect to wall
     collisions, the energy exchange dynamics reduces to the stochastic
     evolution described by equation (\ref{mastereq}).}  
   \label{fig.lb}
\end{figure}

Consider a lattice of confining two-dimensional cells, each containing a
single hard-disc particle, and such that particles in neighbouring cells
may perform elastic collisions with each other, thereby exchanging
energy. Such mechanical systems with hard-core confining mechanisms have
recently been considered in \cite{GG08a, GG08b} (see figure
\ref{fig.lb}). In these systems, one distinguishes the local dynamics from
the interacting dynamics. On the one hand, the local dynamics are
characterized by a wall-collision frequency, $\nu_\mathrm{W}$, which
depends on the geometry of the confining cell as well as on the kinetic
energy of the moving particles. On the other hand, the interacting dynamics
are characterized by the frequency of binary collisions, \emph{i.e.} the
frequency of collisions 
between neighbouring particles, $\nu_\mathrm{B}$. 

Under specific conditions, the scale separation,
$\nu_\mathrm{B}\ll\nu_\mathrm{W}$, is achieved, which is to say that
individual particles typically perform many collisions with the walls of
their confining cells, rattling about their 
cages at higher frequency than that of binary collisions. In this regime,
Liouville's equation governing the time 
evolution of phase-space densities reduces to a master equation for the
time evolution of local energies. Moreover the validity of this reduction
is controlled by the scale separation between the two collision
frequencies $\nu_\mathrm{B}$ and $\nu_\mathrm{W}$, 
and becomes exact in the limit of vanishing binary collision frequency. On
the contrary, in the absence of a clear separation between binary and wall
collision frequencies, the dynamics is not reducible to such a stochastic
description and cannot be addressed in the context of this paper --see
\cite{GG08b} for a discussion of that limit. 

Throughout this article, we assume the validity of the scale separation
between the wall and binary collision frequencies and focus on the stochastic
reduction of the energy exchange dynamics, considering, as our starting
point, the mesoscopic level description of the time evolution of
probability densities as a stochastic evolution. The time evolution is thus
specified by a  master equation which accounts for the energy exchanges
between neighbouring cells. Note that the dimensionality of the dynamics
specifies the maximum dimension of the energy cell array we may
consider. In the case of the dynamics depicted in figure \ref{fig.lb}, the
array of energy variables would be two-dimensional. However, without loss 
of generality, we can simplify the description and consider instead a 
one-dimensional array according to which every cell has two neighbours
instead of four, one on each side\footnote{Though it is of
  course possible to consider a two-dimensional array, as for instance
  shown in figure \ref{fig.lb}, we will focus our attention on
  one-dimensional arrays because, in the presence of a temperature
  gradient along, say, the horizontal direction, no conduction takes place in
  the vertical direction. The only relevant energy exchange processes
  happen along the direction of the temperature gradient.}.
A similar construction can be carried out in three dimensions, starting
with systems of confined hard balls. 

We can thus leave aside the underlying dynamics and consider instead a
system of $N$ cells along a one-dimensional axis with energies 
$\{\epsilon_1,\dots,\epsilon_N\}$ and let $\ple_N(\epsilon_1,\dots,
\epsilon_N, t)$ denote the time-dependent energy distribution associated
with this system. The time evolution of this object is specified according to  
\begin{eqnarray}
  \lo
  \partial_t \ple_N(\epsilon_1,\dots, \epsilon_N, t) = 
  \label{mastereq}\\
  \sum_{a = 1}^N
   \int \ud\eta 
  \Big[W(\epsilon_a + \eta, \epsilon_{a+1} - \eta| \epsilon_a,
  \epsilon_{a+1}) \ple_N(\dots,\epsilon_a + \eta, \epsilon_{a+1} - \eta, 
  \dots, t) \nonumber\\
  \hskip 1.75cm
  - W(\epsilon_a, \epsilon_{a+1} | \epsilon_a - \eta, \epsilon_{a+1} +
  \eta) \ple_N(\dots, \epsilon_a, \epsilon_{a+1}, \dots, t)\Big]\,,
 \nonumber
\end{eqnarray}
where, for the sake of the argument, we assume periodic boundary conditions
and identify cells $a = N+1$ and $a=1$. 

The kernel $W$ specifies the energy transition rates, whereby a pair of
energies $\{\epsilon_a,\epsilon_b\}$ exchange an amount $\eta$ of
energy. In the case of systems such as shown in figure \ref{fig.lb}, it is
given by 
\begin{eqnarray}
\fl W(\epsilon_a, \epsilon_b | \epsilon_a - \eta, \epsilon_b + \eta) = 
\frac{2 \rhom m^2}{(2\pi)^2|\mathcal{L}_{\rho,\rhom}(2)|}
\int\ud\phi\ud \bi{R} 
\int_{\hat{\bi{e}}_{ab}\cdot\bi{v}_{ab} > 0}
\ud \bi{v}_a\ud \bi{v}_b
\label{W}\\
\lo \times \hat{\bi{e}}_{ab}\cdot\bi{v}_{ab}\, 
\delta\left(\epsilon_a - \frac{m}{2}v_a^2\right)
\delta\left(\epsilon_b - \frac{m}{2}v_b^2\right)
 \delta\left(\eta -
  \frac{m}{2}[(\hat{\bi{e}}_{ab}\cdot\bi{v}_a)^2  -  
(\hat{\bi{e}}_{ab}\cdot\bi{v}_b)^2 ]\right)\,,
\nonumber
\end{eqnarray}
where $\hat{\bi{e}}_{ab}$ denotes the unit vector joining particles $a$ and
$b$ with respective velocities $\bi{v}_a$ and $\bi{v}_b$, $\phi$ is the
angle between the direction of this unit vector and a 
reference axis, $\bi{R}$ denotes the position of the center of mass of the
two particles, $m$ their masses, $\rhom$ their radii, and
$|\mathcal{L}_{\rho,\rhom}(2)|$ is the configuration-space volume which
they occupy, with $\rho$ a parameter characterizing the geometry of the
confining cell. Thus $\eta$ is the amount of energy exchanged by the two
particles of respective energies $\epsilon_a$ and $\epsilon_b$ in the
collision process. 

The expression (\ref{W}) of $W$ extends beyond the mechanical systems
described above. It applies to all mechanical models in which locally
confined particles interact with their nearest neighbours through hard-core
collisions. The dimensionality of the dynamics, whether two or three, is of
course relevant to the specific form of the kernel, in particular as far
the computation of the velocity integrals goes. However, whether the
underlying dynamics is two- or three-dimensional, it can be shown that the
transport properties of the energy exchange process to be established
below are obtained in similar ways.

In the above expression, one shows that the spatial integral $\int \ud \phi
\ud \bi{R}$ decouples from the velocity integral
$\int_{\hat{\bi{e}}_{ab}\cdot\bi{v}_{ab} > 0} \ud \bi{v}_a\ud \bi{v}_b$. 
The former quantity represents the volume that the center of mass of
particles $a$ and $b$ can occupy given that the two particles are in
contact, performing a collision. Now rescaling the time variable by a
factor $2\rhom \int\ud\phi\ud \bi{R}/[\sqrt{\pi m}|
\mathcal{L}_{\rho,\rhom}(2)|]$, which amounts to converting time to the
units of the inverse of the square root of an energy, we can rid equation
(\ref{W}) of all geometric factors and write the expression of the kernel
in terms of Jacobi elliptic functions of the first kind, denoted
$K$. Assuming $\epsilon_a\le\epsilon_b$, the kernel takes the expression  
\begin{equation}
  \label{kernel}
  \fl 
  W(\epsilon_a, \epsilon_b | \epsilon_a - \eta, \epsilon_b + \eta) =
    \sqrt{\frac{2}{\pi^3}} \times
    \left\{
      \begin{array}{l@{\quad}l}
        \sqrt{\frac{1}{\epsilon_a}}
        K \left(\frac{\epsilon_b + \eta}{\epsilon_a}\right)\,,&
        -\epsilon_b < \eta < \epsilon_a - \epsilon_b\,,\\
        \sqrt{\frac{1}{\epsilon_b + \eta}}
        K \left(\frac{\epsilon_a}{\epsilon_b + \eta}\right)\,,&
        \epsilon_a - \epsilon_b < \eta<0\,,\\
        \sqrt{\frac{1}{\epsilon_b}}
        K \left(\frac{\epsilon_a - \eta}{\epsilon_b}\right)\,,&
        0<\eta<\epsilon_a\,.
      \end{array}
    \right.
\end{equation}
For $\epsilon_b\le\epsilon_a$, the kernel is defined in a similar way,
using the symmetry of $W$ with respect to its arguments, \emph{i.e.}
exchanging $\epsilon_a$ and $\epsilon_b$ and changing the sign of $\eta$.

The rate at which two neighbouring cells with energies $\epsilon_a$ and
$\epsilon_b$ exchange energy is obtained by integrating $W$ over the range
of possible values of the amount $\eta$ of energy exchanged,
\begin{equation}
\nu(\epsilon_a, \epsilon_b) = \int_{-\epsilon_b}^{\epsilon_a} \ud \eta\,
W(\epsilon_a, \epsilon_b | \epsilon_a - \eta, \epsilon_b + \eta)\,,
 \label{collfrequencydef}
\end{equation}
which, using equation (\ref{kernel}), is easily computed in terms
of first and second kind elliptic functions, $K$ and $E$, to read
\begin{equation}
  \nu(\epsilon_a, \epsilon_b) = \sqrt{\frac{8\epsilon_b}{\pi^3}}
  \left[ 2 E\left(\frac{\epsilon_a}{\epsilon_b}\right)
    - \left(1 - \frac{\epsilon_a}{\epsilon_b}\right)
    K\left(\frac{\epsilon_a}{\epsilon_b}
    \right) \right] \quad (\epsilon_b \ge \epsilon_a)\,.
 \label{collfrequency}
\end{equation}
This is a symmetric function of $\epsilon_a$, $\epsilon_b$, \emph{i.e.}   
$\nu(\epsilon_a, \epsilon_b) =   \nu(\epsilon_b, \epsilon_a)$ if
$\epsilon_a > \epsilon_b$. It is homogeneous in the sense that, given a
positive constant $\alpha$, we have $\nu(\alpha \epsilon_a, \alpha
\epsilon_b) = \sqrt{\alpha}\nu(\epsilon_a, 
\epsilon_b)$. 

Likewise, the average heat exchanged between two cells at respective
energies $\epsilon_a$ and $\epsilon_b$, is
\begin{eqnarray}
    j(\epsilon_a,\epsilon_b) &\equiv& \int_{-\epsilon_b}^{\epsilon_a} \ud
    \eta \, \eta\,  
    W(\epsilon_a, \epsilon_b| \epsilon_a - \eta, \epsilon_b + \eta)\,,
    \nonumber\\
    &=& \frac{2}{3}(\epsilon_a - \epsilon_b) \nu(\epsilon_a, \epsilon_b)\,,
    \label{heatcurrent}
     \\
     &=& \frac{2(\epsilon_a - \epsilon_b)}{3}
     \sqrt{\frac{8\epsilon_b}{\pi^3}}
     \left[ 2 E\left(\frac{\epsilon_a}{\epsilon_b}\right)
       - \left(1 - \frac{\epsilon_a}{\epsilon_b}\right)
       K\left(\frac{\epsilon_a}{\epsilon_b}\right)
       \right] \quad (\epsilon_b \ge \epsilon_a)\,,
     \nonumber  
\end{eqnarray}
which is an antisymmetric function, {\em i.e.} defined by $j(\epsilon_a,
\epsilon_b) = -j(\epsilon_b, \epsilon_a)$ when $\epsilon_a > \epsilon_b$.

\section{\label{sec.ss}Stationary states}

A stationary state of (\ref{mastereq}) is a time-independent
probability distribution  $\pss_N(\epsilon_1,\dots,\epsilon_N)$ such that,
for all pairs $(\epsilon_a,\epsilon_b)$, 
\begin{eqnarray}
  \int \ud\eta 
  \Big[W(\epsilon_a + \eta, \epsilon_b - \eta| \epsilon_a,
  \epsilon_b) \pss_N(\dots,\epsilon_a + \eta, \dots, \epsilon_b - \eta,
  \dots) \nonumber\\
  \hskip 1.cm
  - W(\epsilon_a, \epsilon_b | \epsilon_a - \eta, \epsilon_b + \eta)
  \pss_N(\dots, \epsilon_a, \dots, \epsilon_b, \dots)\Big] = 0\,.
\label{ssmastereq}
\end{eqnarray}

\subsection{Equilibrium states}

Equilibrium measures have densities which depend on the local energies
through their sum only, {\em e.g.} the micro-canonical distribution
$\delta(\epsilon_1 + \dots + \epsilon_N - E)$ and the canonical
distribution $\beta^N \exp[-\beta(\epsilon_1+\dots+\epsilon_N)]$. The
former is associated with an isolated $N$-cells system, the latter to a
system in contact with a thermal bath at fixed inverse temperature
$\beta = 1/T$.  Note that we assume the temperature
is measured in the units of the energy, or, equivalently, that the
Boltzmann constant is unity, $\kb \equiv 1$.

The equilibrium average of the rate of energy exchanges
(\ref{collfrequencydef}) is the collision frequency, which in the canonical
ensemble at temperature $T$, is given by  
\begin{equation}
  \nub = \frac{1}{T^2}\int \ud \epsilon_a \ud \epsilon_b
  \nu(\epsilon_a,\epsilon_b)  
  \exp[-(\epsilon_a + \epsilon_b)/ T] = \sqrt{T}\,.
  \label{nub}
\end{equation}
Hence the choice of the time scale.

We further notice that, for finite size systems of $N$ cells
and total energy $E = N T$, {\em i.e.} average energy $T$ per cell, the
collision frequency is obtained from the micro-canonical average, 
\begin{eqnarray}
  \lo \nub &= \frac{1}{(NT)^2} \int_0^{NT} \ud \epsilon_a \int_0^{NT -
    \epsilon_a} \ud 
  \epsilon_b \nu(\epsilon_a,\epsilon_b)  (N-1)(N-2) 
  \bigg(1-\frac{\epsilon_a + \epsilon_b}
  {N T}\bigg)^{N-3}\,,\nonumber\\
  &= \sqrt{T} + \mathcal{O}(N^{-1})\,.
  \label{nubN}
\end{eqnarray}

For future sake, we also notice the identity 
\begin{equation}
  \int \ud x \ud y (x - y)^2 \nu(x, y) \exp(-x - y) = 3\,.
  \label{esqnub}
\end{equation}

\subsection{Non-equilibrium states}

A non-equilibrium stationary state occurs {\em e.g.} when the two ends of a
one-dimensional channel are put in contact with thermal baths at inverse
temperatures $\beta_-$ and $\beta_+$, with $\beta_- \ne \beta_+$, so that a
stationary heat flux will flow from the hot to the cold end. 
One might hope that the corresponding stationary state would have the local
product structure 
\begin{equation}
  \pss_N(\epsilon_1,\dots,\epsilon_N)
  = \beta_1\cdots\beta_N
  \exp[-\beta_1 \epsilon_1 - \dots - \beta_N \epsilon_N]\,,
  \label{ness}
\end{equation}
with a temperature profile specified according to Fourier's law
$\partial_x [\kappa(x) \partial_x T(x)] = 0$. However such a distribution
is not stationary since an energy exchange displaces an amount of energy
between two cells at different local temperatures. We will however see in
Section \ref{sec.ce} that (\ref{ness}) is indeed a good approximation to
the actual non-equilibrium stationary state and can be used to compute the
linear response.

\section{\label{sec.en}Conservation of energy}

In order to analyze heat transport in our system, we need to consider the
time evolution of local temperatures 
\begin{equation}
  \kb T(a, t) \equiv \langle \epsilon_a \rangle\,.
  \label{deftemp}
\end{equation}
We have, using
equation (\ref{mastereq}), and assuming we have a one-dimensional
lattice of cells so that only terms $(a-1,a)$ and $(a,a+1)$ contribute,
\begin{eqnarray}
  \fl \partial_t \kb T(a, t) = \int \ud \epsilon_1\dots \ud
  \epsilon_N \epsilon_a \partial_t \ple_N(\epsilon_1,\dots, \epsilon_N, t)\,,
  \nonumber\\ 
  \fl = \frac{1}{2}\sum_{a',b'}\int \ud \epsilon_1\dots \ud \epsilon_N \ud
  \eta \Big[\epsilon_a W(\epsilon_{a'} + \eta, \epsilon_{b'} - \eta| 
  \epsilon_{a'}, \epsilon_{b'}) 
  \ple_N(\dots,\epsilon{_a'} + \eta, \dots, \epsilon_{b'} - \eta,
  \dots, t) \nonumber\\
  - \epsilon_a W(\epsilon_{a'}, \epsilon_{b'} | \epsilon_{a'} - \eta, 
  \epsilon_{b'} + \eta) 
  \ple_N(\dots, \epsilon_{a'}, \dots, \epsilon_{b'}, \dots, t)\Big]\,,
  \nonumber\\ 
  \fl=\int \ud \epsilon_1\dots \ud \epsilon_N \ud
  \eta \Big[\epsilon_a W(\epsilon_a + \eta, \epsilon_{a+1} - \eta| \epsilon_a,
  \epsilon_{a+1}) \ple_N(\dots,\epsilon_a + \eta, \epsilon_{a+1} - \eta,
  \dots, t) \nonumber\\
  - \epsilon_a W(\epsilon_a, \epsilon_{a+1} | \epsilon_a - \eta,
  \epsilon_{a+1} + \eta) \ple_N(\dots, \epsilon_a, \epsilon_{a+1}, \dots,
  t)\nonumber\\
  + \epsilon_a W(\epsilon_{a-1} + \eta, \epsilon_{a} - \eta| \epsilon_{a-1},
  \epsilon_{a}) \ple_N(\dots,\epsilon_{a-1} + \eta, \epsilon_{a} - \eta,
  \dots, t) \nonumber\\
  - \epsilon_a W(\epsilon_{a-1}, \epsilon_{a} | \epsilon_{a-1} - \eta,
  \epsilon_{a} + \eta) \ple_N(\dots, \epsilon_{a-1}, \epsilon_{a}, \dots,
  t) \Big]\,, \nonumber\\ 
  \fl=\int \ud \epsilon_1\dots \ud \epsilon_N \ud
  \eta \Big[-\eta W(\epsilon_a, \epsilon_{a+1}| \epsilon_a  - \eta,
  \epsilon_{a+1} + \eta) \ple_N(\dots,\epsilon_a, \epsilon_{a+1},
  \dots, t) \nonumber\\
  + \eta W(\epsilon_{a-1}, \epsilon_{a} | \epsilon_{a-1} - \eta,
  \epsilon_{a} + \eta) \ple_N(\dots, \epsilon_{a-1}, \epsilon_{a}, \dots,
  t)\Big]\,.
\end{eqnarray}
The two terms on the RHS of the last line denote the energy currents
flowing between the neighbouring cells,
\begin{equation}
  \partial_t \kb T(a, t) = J_{a-1,a}(t)  - J_{a,a+1}(t)\,,
  \label{energyconservation}
\end{equation}
where 
\begin{equation}
 J_{a,b}(t) \equiv \int \ud \epsilon_1\dots \ud \epsilon_N 
  j(\epsilon_a,\epsilon_{b}) \ple_N(\epsilon_1, \dots, \epsilon_{N}, t)
  \label{current}
\end{equation}
denotes the average energy flux from cell $a$ to cell $b$ with respect to
the distribution $\ple_N$. Thus equation
(\ref{energyconservation}) is a conservation equation for the 
energy. We remark here that we have dropped the dimensional factors,
setting the cells lengths to unity.

Fourier's law relates the heat current to the local gradient of temperature
through a linear law which involves the coefficient of heat
conductivity. We now set out to derive Fourier's law and thereby compute
this quantity.

\section{\label{sec.helf}Helfand moment}

The heat conductivity which is associated with heat transport can be computed
by considering the linear growth in time of the mean squared Helfand
moment for that process, which undergoes a deterministic diffusion in phase
space \cite{Hel60}. This is a generalisation of Einstein's formula which
relates the diffusion coefficient to the mean squared displacement. 

Assuming a one-dimensional lattice of cells with unit distance between
neighbouring cells, we may write the Helfand moment associated with heat
transport as
\begin{equation}
  H(t) = \sum_{a = 1}^N a \epsilon_a(t)\,.
  \label{defhelf}
\end{equation}
Changes in this quantity occur whenever an energy exchange takes place
between any two neighbouring cells. Thus considering a stochastic
realisation of the energy exchange process, we have a sequence of times
$\tau_n$ at which a given pair of cells $(k_n, k_n+1)$ performs an energy
exchange $\eta(\epsilon_{k_n}, \epsilon_{k_n+1})$, {\em i.e.}
$\epsilon_{k_n} \to \epsilon_{k_n} - \eta$ and $\epsilon_{k_n+1} \to
\epsilon_{k_n+1} + \eta$. The corresponding Helfand moment therefore
evolves according to  
\begin{equation}
  H(\tau_n) = H(\tau_{n-1}) + \eta(\epsilon_{k_n}, \epsilon_{k_n+1})\,,
  \label{helfn}
\end{equation}
and has overall displacement
\begin{equation}
  \Delta H(\tau_n) \equiv H(\tau_n) - H(\tau_0) = 
  \sum_{i=1}^n \eta(\epsilon_{k_i}, \epsilon_{k_i+1})\,.
  \label{deltahelfn}
\end{equation}

The thermal conductivity can be computed in terms of the equilibrium
average of the mean squared displacement of the Helfand moment according to
\begin{equation}
  \kappa = \lim_{N\to\infty} \frac{1}{N (E/N)^2} 
  \lim_{n\to\infty} 
  \bigg \langle \frac{1}{2\tau_n}  \Delta H(\tau_n)^2 \bigg\rangle_{E/N}\,,
\label{kappahelf}
\end{equation}
where $\langle . \rangle_{E/N}$ denotes a micro-canonical equilibrium
average of the $N$ cells system at energy $E \equiv NT$, which involves 
both energy exchanges and relaxation times between then.
When $N$ 
is large, boils down to an average with respect to the canonical
distribution at temperature $T$. In equation (\ref{kappahelf}), the time
$\tau_n$ corresponding to $n$ events is, by the law of large numbers, the
typical time it takes for $n$ collision events to occur in a system of $N$
cells, which is easily expressed in terms of the collision frequency
(\ref{nub}), 
\begin{equation}
  \lim_{n\to\infty} \frac{n}{\tau_n} = N \nub\,.
\label{taun}
\end{equation}

Substituting $\Delta H$ from equation (\ref{deltahelfn}) into equation
(\ref{kappahelf}), we obtain the thermal conductivity,
\begin{eqnarray}
  \kappa = \lim_{N\to\infty} \frac{1}{N T^2} 
  \lim_{n\to\infty} 
  \Bigg[\sum_{i=1}^n \bigg \langle \frac{1}{2\tau_n} 
    \eta(\epsilon_{k_i}, \epsilon_{k_i+1})^2
    \bigg \rangle_{T} \label{kappa1}\\
    \hskip 4cm
    + 2 \sum_{i=1}^{n-1} \sum_{j = i+1}^n
    \bigg \langle \frac{1}{2\tau_n} 
    \eta(\epsilon_{k_i}, \epsilon_{k_i+1}) \eta(\epsilon_{k_j},
    \epsilon_{k_j+1}) \bigg\rangle_{T}
  \Bigg]\,,
  \nonumber
\end{eqnarray}
which is equivalent to the corresponding result using the Green-Kubo formula.
 
The first of the two terms in the brackets on the RHS of (\ref{kappa1}) is
a sum of $n$ identical static averages, which, for large $N$, is
approximated by 
the canonical equilibrium average
\begin{eqnarray}
  \fl \bigg \langle \eta(\epsilon_a, \epsilon_b)^2 \bigg \rangle_T 
   = \frac{1}{\nub T^2}
   \int \ud \epsilon_a  \ud \epsilon_b\, \ud \eta\, \eta^2 
   W(\epsilon_a, \epsilon_b | \epsilon_a - \eta, \epsilon_b + \eta) 
  \exp[-(\epsilon_a + \epsilon_b)/T]\,,
  \nonumber\\
  = 2 T^{2}\,.
  \label{autocorrel}
\end{eqnarray}
Notice here that, in order to compute the second moment of the average
energy exchanged between the two cells, we divided $W$ by $\nub$, the rate
at which these exchanges occur.

The second term on the RHS of equation (\ref{kappa1}), on the other hand,
is a sum of dynamic averages which are generally difficult to compute. 
We contend that it goes to zero when $N\to\infty$. This is to say that
cross-correlations of energy transfers $\langle \eta(\epsilon_{k_i},
\epsilon_{k_i+1}) \eta(\epsilon_{k_j}, \epsilon_{k_j+1}) \rangle_{T}$ exist
only so long as the system size is finite. The reason for this 
is that, in large enough systems, new
energies keep entering the dynamic averages in equation (\ref{kappa1}),
as if the systems were in contact with stochastic reservoirs, which is
enough to destroy these correlations. 

In the following section, we turn to the evaluation of the non-equilibrium
stationary state and will provide more definitive arguments that concur
with these heuristic reasoning. In section \ref{sec.nr} we discuss the
results of numerical computations of the Helfand moment which provide
further insight into the decay of these dynamic averages as $N\to\infty$.

Accepting the claim that dynamic averages vanish in equation
(\ref{kappa1}) and thus that only static averages contribute to the heat
conductivity, we conclude that the thermal conductivity associated with the
process defined by (\ref{mastereq}) is equal to the collision
frequency\footnote{The dimensions of these quantities differ by a length
  squared --the separation between the energy cells-- which we have set to
  unity.}, which, for a  general equilibrium temperature $T$, is
\begin{equation}
\kappa = \nub = \sqrt{T}\,.
\label{kappanub}
\end{equation}

\section{\label{sec.ce} Chapman-Enskog gradient expansion}

Consider the marginals of the $N$-cell distribution function, given by
the one- and two-cell distribution functions,
\begin{equation}
  \eqalign{
    \Pone_a(\epsilon_a, t) = \int \prod_{i\ne a}\ud\epsilon_i
    \PN(\epsilon_1, \dots, \dots, \epsilon_N, t)\,,\\
    \Ptwo_{a,b}(\epsilon_a, \epsilon_b, t) = \int \prod_{i\ne a,b}
    \ud\epsilon_i \PN(\epsilon_1, \dots, \dots, \epsilon_N, t)\,.
  }
\end{equation}
The time evolution of the former can be written in terms of the latter,
\begin{eqnarray}
  \partial_t \Pone_a(\epsilon, t) &=& \int \prod_{i\ne a}\ud\epsilon_i
  \partial_t \PN(\epsilon_1, \dots, \epsilon_{a-1}, \epsilon, \epsilon_{a+1}, 
  \dots, \epsilon_N, t)\,,\nonumber\\
  &=& \int \ud\epsilon' \ud\eta 
  \Big[W(\epsilon + \eta, \epsilon' - \eta| \epsilon, \epsilon')
  \Ptwo_{a,a+1}(\epsilon + \eta, \epsilon' - \eta)
  \nonumber\\
  &&- W( \epsilon, \epsilon'|\epsilon - \eta, \epsilon' + \eta)
  \Ptwo_{a,a+1}(\epsilon, \epsilon')
  \nonumber\\
  &&+W(\epsilon + \eta, \epsilon' - \eta| \epsilon, \epsilon')
  \Ptwo_{a,a-1}(\epsilon + \eta, \epsilon' - \eta)
  \nonumber\\
  &&- W( \epsilon, \epsilon'|\epsilon - \eta, \epsilon' + \eta)
  \Ptwo_{a,a-1}(\epsilon, \epsilon')\Big]\, .
  \label{1Pevol}
\end{eqnarray}
This equation is the first step of the BBGKY hierarchy.
We want to solve this equation in the non-equilibrium stationary state
which is generated by putting the system boundaries in contact with thermal
reservoirs at two different temperatures, $T_- \ne T_+$, with respective
distributions $P_L(\epsilon) = \beta_-\exp(-\beta_-\epsilon)$ and
$P_R(\epsilon) = \beta_+\exp(-\beta_+\epsilon)$, at the corresponding
inverse temperatures $\beta_-$ and $\beta_+$. 

In the presence of such a temperature gradient, we expect that a uniform
stationary heat current, as defined by equation (\ref{current}) and
henceforth denoted $\JH$, will establish itself throughout the
system. Provided the local temperature gradient $(T_+ - T_-)/N$ is small,
this current would be given, according to Fourier's law, by the product of
the local temperature gradient and heat conductivity $\kappa$:
\begin{equation}
  \JH = -  \kappa\,(T_{a+1}- T_a) + \mathcal{O}[(T_+- T_-)/N]^2\,.
  \label{fourierlaw}
\end{equation}
This equality should hold for any $0\leq a\leq N$, $a=0$ and $a = N+1$
corresponding to the two thermal reservoirs. Notice that the heat
conductivity is itself a function of the temperature and thus varies from
site to site. 

In order to derive Fourier's law (\ref{fourierlaw}), as well as the
expression of the heat conductivity $\kappa$, and thus establish  equation
(\ref{kappanub}), we compute the heat current using equation
(\ref{current}), which, in terms of the 
two-cell distribution $\Ptwo_{a,a+1}(\epsilon, \epsilon')$, becomes 
\begin{equation}
  \JH \equiv \int \ud \epsilon \ud \epsilon'
  j(\epsilon,\epsilon') \Ptwo_{a,a+1}(\epsilon, \epsilon')\,.
  \label{currenttwop}
\end{equation}
As noticed above, $\JH$ on the LHS of this equation is defined to first
order in the local temperature gradients. We therefore need to compute 
$\Ptwo_{a,a+1}(\epsilon, \epsilon')$ to that order as well, which is to say
it must be a stationary state of equation (\ref{1Pevol}) up to second order
in the local temperature gradient.

In order to find this approximate stationary state, we perform a cluster
expansion of the two cell distribution function and start by assuming
that it factorises into the product of two one-cell distribution functions,
\begin{equation}
  \Ptwo_{a,b}(\epsilon, \epsilon')  = 
  \Pone_{a}(\epsilon) \Pone_b(\epsilon') \, .
  \label{factorPtwo}
\end{equation}
This solution is as yet incomplete and must be understood as being only the
first step in obtaining the consistent solution which is to be written down
below.

Substituting equation (\ref{factorPtwo}) into equation (\ref{1Pevol}), it
thus reduces to a Boltzmann-Kac equation, which, 
owing to the symmetries of the kernel, can be written as 
\begin{eqnarray}
  \partial_t \Pone_a(\epsilon,t)
  &= \int \ud\epsilon' \ud\eta 
  W(\epsilon, \epsilon'|\epsilon - \eta, \epsilon' + \eta) 
  \Big[\Pone_{a}(\epsilon + \eta) \Pone_{a+1}(\epsilon' - \eta) 
    \label{Kaceq}\\
  &- \Pone_{a}(\epsilon) \Pone_{a+1}(\epsilon')
  + \Pone_{a}(\epsilon + \eta) \Pone_{a-1}(\epsilon' - \eta) 
  - \Pone_{a}(\epsilon) \Pone_{a-1}(\epsilon')\Big] \, .
  \nonumber
\end{eqnarray}

A first approximation to the stationary state of this equation is
given by the local thermal equilibrium solution,
\begin{equation}
  \Pone_{a}(\epsilon) = \beta_{a} e^{-\beta_{a}\epsilon}\,.
  \label{ssP}
\end{equation}
By definition of the local temperature, we have
\begin{equation}
  \int_0^\infty \ud\epsilon \, \epsilon \Pone_a(\epsilon) = \beta_a^{-1} 
  = T_a\,.
\end{equation}
One verifies that the form (\ref{ssP}) is an approximate
stationary solution of the Boltzmann-Kac equation (\ref{Kaceq}).
Indeed, let $\beta_a = \beta$ denote the equilibrium temperature,
$\beta_{a\pm 1} = \beta \pm \delta\beta$. We have 
\begin{eqnarray}
  \fl
  \Pone_{a}(\epsilon + \eta) \Pone_{a+1}(\epsilon' - \eta) 
  - \Pone_{a}(\epsilon) \Pone_{a+1}(\epsilon') = 
  \beta_a\beta_{a+1} e^{-\beta_a\epsilon-\beta_{a+1}\epsilon'}
  [e^{\eta(\beta_{a+1}-\beta_a)} - 1]\,,  \nonumber\\
  = \beta^2e^{-\beta(\epsilon + \epsilon')}\eta \delta\beta
  + \mathcal{O}(\delta\beta^2)\, ,
  \label{sum1}
\end{eqnarray}
Likewise 
\begin{equation}
  \Pone_{a}(\epsilon + \eta) \Pone_{a-1}(\epsilon' - \eta) 
  - \Pone_{a}(\epsilon) \Pone_{a-1}(\epsilon') 
  = - \beta^2e^{-\beta(\epsilon + \epsilon')}\eta \delta\beta
  + \mathcal{O}(\delta\beta^2)\, .
  \label{sum2}
\end{equation}
Therefore, to $\mathcal{O}(\delta\beta^2)$, the sum of equations
(\ref{sum1}) and (\ref{sum2}) vanishes. This is to say the local thermal
equilibrium  distributions (\ref{ssP}) give approximate stationary states
of the Boltzmann-Kac equation (\ref{Kaceq}) to that order.

Going back to the two-cell distribution, equation (\ref{factorPtwo}), we
proceed to compute the next order of the BBGKY hierarchy. One might hope
that local equilibrium solutions are also a good approximation to the
two-cell distribution function. This is however not the case. We can see
this by considering the time evolution of the two-cell distribution,
\begin{eqnarray}
  \fl 
  \partial_t \Ptwo_{a,a+1}(\epsilon, \epsilon')
  = \int \ud\eta 
  W(\epsilon, \epsilon'|\epsilon - \eta, \epsilon' + \eta) 
  \Big[\Ptwo_{a,a+1}(\epsilon + \eta, \epsilon' - \eta) 
  -\Ptwo_{a,a+1}(\epsilon, \epsilon')\Big]
  \label{twopevol}\\
  \fl \hskip .7cm
  + \int \ud\epsilon'' \ud\eta 
  W(\epsilon, \epsilon''|\epsilon - \eta, \epsilon'' + \eta) 
  \Big[\Pthree_{a-1,a,a+1}(\epsilon'' - \eta, \epsilon + \eta, \epsilon')
  -\Pthree_{a-1,a,a+1}(\epsilon'', \epsilon, \epsilon')\Big]
  \nonumber\\
  \fl \hskip .7cm
  + \int \ud\epsilon'' \ud\eta 
  W(\epsilon', \epsilon''|\epsilon' - \eta, \epsilon'' + \eta) 
  \Big[\Pthree_{a,a+1,a+2}(\epsilon, \epsilon' + \eta,\epsilon'' - \eta)
  -\Pthree_{a,a+1,a+2}(\epsilon, \epsilon',\epsilon'')\Big]\,.
  \nonumber
\end{eqnarray}
We momentarily suppose that the three-cell distributions can be
consistently written as a product of one-cell distributions,
\begin{equation}
  \Pthree_{a,b,c}(\epsilon, \epsilon', \epsilon'')  
  = \Pone_{a}(\epsilon) \Pone_b(\epsilon')  \Pone_c(\epsilon'') \, .
  \label{factorPthree}
\end{equation}
Plugging this form into equation (\ref{twopevol}) and assuming a stationary
state, we arrive, after expanding all terms to $\mathcal{O}(\delta\beta)$,
to the equation 
\begin{equation}
  \fl 0 = \beta^2 e^{-\beta(\epsilon + \epsilon')} \delta\beta\bigg\{
  j(\epsilon, \epsilon') 
  - \beta \int \ud \epsilon'' e^{-\beta \epsilon''} j(\epsilon, \epsilon'')
  + \beta \int \ud \epsilon'' e^{-\beta \epsilon''} j(\epsilon',
  \epsilon'') 
  \bigg\}\,.
  \label{iscurrentgradient}
\end{equation}
This is however a contradiction since the current does not have the
gradient form which (\ref{iscurrentgradient}) implies,
\begin{equation}
  j(\epsilon, \epsilon') \ne 
   \beta \int \ud \epsilon'' e^{-\beta \epsilon''} j(\epsilon, \epsilon'')
  - \beta \int \ud \epsilon'' e^{-\beta \epsilon''} j(\epsilon',
  \epsilon'')\,. 
  \label{currentnotgradient}
\end{equation}
Indeed, the only possible solution of this equation is a current given in
terms of the difference of two local functions, say $j(\epsilon,\epsilon')
= f(\epsilon) - f(\epsilon')$, which, clearly, equation (\ref{heatcurrent})
does not satisfy.

To work around this difficulty, we must go back to equation
(\ref{factorPtwo}) and perform a cluster expansion to include
$\mathcal{O}(\delta \beta)$ corrections in the form
\begin{eqnarray}
  \Ptwo_{a,a\pm 1}(\epsilon, \epsilon')  &= 
  \Pone_{a}(\epsilon) \Pone_{a\pm1}(\epsilon') 
  + (\beta_{a\pm1}-\beta_a) \Qtwo_{a,a\pm 1} (\epsilon, \epsilon')\,,
  \nonumber\\
  &\equiv
  \beta^2 e^{-\beta(\epsilon + \epsilon')}
  \left[1 \mp \delta \beta \epsilon' \pm \frac{\delta \beta}{\beta}
  \qtwo_{a,a\pm1}(\beta\epsilon, \beta\epsilon')\right]\,,
  \label{factorPtwoagain}
\end{eqnarray}
where the first two terms in the second line come from the expansion of
the one-cell distributions and $\qtwo$ is one plus an expression derived
from $\Qtwo$. 

Notice that, for definiteness, we require that $\int \ud \epsilon'
\Ptwo_{a,b}(\epsilon, \epsilon')  = \Pone_a(\epsilon)$, implying
\begin{equation}
  \int \ud \epsilon' \Qtwo_{a,b} (\epsilon, \epsilon') = 0\,.
  \label{qorth}
\end{equation}

Substituting the form (\ref{factorPtwoagain}) into the RHS of equation
(\ref{1Pevol}), we must have the cancellation of all
$\mathcal{O}(\delta\beta)$ terms, and thus 
\begin{eqnarray}
  \lo 
  0 = \int \ud\epsilon' \ud\eta e^{-\beta \epsilon'}
  W(\epsilon, \epsilon'|\epsilon - \eta, \epsilon' + \eta) 
  \Big\{\qtwo_{a,a+1}[\beta(\epsilon + \eta)\,, \beta(\epsilon' -
    \eta)] 
  \label{twopointKaceq}
  \\
  \hskip 1cm
  - \qtwo_{a,a+1}(\beta\epsilon, \beta\epsilon')
  - \qtwo_{a,a-1}[\beta(\epsilon + \eta)\,, \beta(\epsilon' - \eta)]
  + \qtwo_{a,a-1}(\beta\epsilon, \beta\epsilon')\Big\}\,.  \nonumber
\end{eqnarray}
Given that the system is large, we expect $\qtwo_{a,a\pm1}$ to converge to
the forms
\begin{equation}
  \eqalign{
    \qtwo_{a,a+1}(x, y) = q(x, y) + \mathcal{O}(\delta \beta)\,,\\
    \qtwo_{a,a-1}(x, y) = q(y, x) + \mathcal{O}(\delta \beta)\,.
  }
\end{equation}
If so, let us consider $q$ to be the sum of a symmetric function $g$ and an
antisymmetric function $h$, $q(x,y) = g(x,y) + h(x,y)$, with $g(x,y) =
g(y,x)$ and $h(x,y) = -h(y,x)$. It is obvious that 
the terms between the brackets of equation (\ref{twopointKaceq}) cancel 
each other with respect to the symmetric part $g$. As far as the
antisymmetric  part, on the other hand, we must have
\begin{equation}
  \fl \int_0^\infty \ud y e^{-y} 
  \left[\int_{-y}^x \ud z W(x,y|x-z,y+z) h(x+z, y - z) - \nu(x,y) h(x,y)
    \right] = 0\,.
  \label{qasym}
\end{equation}
We argue that the only possible solution of equation (\ref{qasym}) is a 
function $h$ which depends on the sum of its arguments, which is obviously
not an antisymmetric function. Therefore $q$ is a  \emph{symmetric}
function of its arguments,
\begin{equation}
  q(y, x) = q(x, y)\,.
  \label{symqtwo}
\end{equation}

With the two-cell distribution (\ref{factorPtwoagain}), we write the
three-cell distribution, now including the second order terms of the cluster
expansion, as 
\begin{eqnarray}
  \fl
  \Pthree_{a-1,a,a+1}(\epsilon, \epsilon', \epsilon'')  = 
  \Pone_{a-1}(\epsilon) \Pone_a(\epsilon')  \Pone_{a+1}(\epsilon'') 
   \nonumber\\
   + \delta \beta \left[\Pone_{a-1}(\epsilon)
     \Qtwo_{a,a+1}(\epsilon', \epsilon'')
     + \Pone_{a+1}(\epsilon'')
      \Qtwo_{a-1,a}(\epsilon, \epsilon') \right]\,,
  \nonumber\\
  = \beta^3e^{-\beta(\epsilon + \epsilon' + \epsilon'')}
  \Bigg\{1 + \delta\beta (\epsilon - \epsilon'') + 
    \frac{\delta \beta}{\beta}
    \left[q(\beta\epsilon, \beta\epsilon') +
    q(\beta\epsilon', \beta\epsilon'')\right]\Bigg\}\,.
\end{eqnarray}
Plugging this expression into equation (\ref{twopevol}), we obtain an
equation containing the terms of equation (\ref{iscurrentgradient}), but
this time with additional terms involving $q$'s. Though this equation is
complicated and yields \emph{a priori} no simple explicit solution for $q$,
it is consistent with the fact that $j$ does not have the gradient
property. 

Nonetheless, finding the exact expression of $q$ is not necessary for our
sake as the symmetry of $q$, equation (\ref{symqtwo}), turns out to be
enough information on the two-cell distribution function to compute
the stationary heat current (\ref{currenttwop}). Indeed, we may write the
stationary current as the average
\begin{equation}
  \JH =\int \ud x \ud y e^{-(x+y)}\left\{1 - \frac{\delta\beta}{\beta}
    [y - q(x,y)]\right\} 
  j\left(\frac{x}{\beta},\frac{y}{\beta}\right)\,.
  \label{sscurrent}
\end{equation}
Notice from equation (\ref{heatcurrent}) that $j(x/\beta,y/\beta) =
\beta^{-3/2}j(x,y)$. Thus, to $\mathcal{O}(\delta\beta)$, there are
only two possible contributions to the heat current that need to be taken
into account. Namely,
\begin{equation}
  \eqalign{
    \int  \ud x \ud y e^{-(x+y)} y j(x, y)\,,\\
    \int  \ud x \ud y e^{-(x+y)} q(x, y) j(x, y)\,.
    }
    \label{contribj}
\end{equation}
However the second of these expressions identically vanishes since $j$ is
an anti-symmetric function of its arguments and $q$, as we saw in equation
(\ref{symqtwo}), is symmetric. The steady state current thus becomes 
\begin{equation}
  \JH = - \frac{\delta\beta}{\beta^{5/2}}
  \int \ud x \ud y e^{-(x+y)} y j(x, y)\,,
\end{equation}
which, by symmetry, can be rewritten as 
\begin{eqnarray}
  \JH =&  \frac{1}{2}\frac{\delta\beta}{\beta^{5/2}}
  \int \ud x \ud y e^{-(x+y)} (x-y) j(x, y)\,,\nonumber\\
  &=  \frac{1}{3}\frac{\delta\beta}{\beta^{5/2}}
  \int \ud x \ud y e^{-(x+y)} (x-y)^2 \nu(x, y)\,,\nonumber\\
  &=  \frac{\delta\beta}{\beta^{5/2}}\,,\nonumber\\
  &= - \sqrt{T}\delta T.
  \label{currentdeltaT}
\end{eqnarray}
where we consecutively used equations (\ref{heatcurrent}) in the second
line, (\ref{esqnub}) in the third one, and the identity $\delta\beta =
-\delta T/T^2$ in the last one. Equation (\ref{currentdeltaT}) is nothing
but Fourier's law (\ref{fourierlaw}), with conductivity
\begin{equation}
  \kappa = \sqrt{T}\,,
  \label{kappaness}
\end{equation}
confirming equation (\ref{kappanub}). Thus, even though there are
$\mathcal{O}(\delta\beta)$ contributions to the stationary state which
arise from the two-point distribution function, the only actual
contribution to heat transport arises from the average with respect to the
local equilibrium part. This result also confirms our claim that the
cross-correlations on the RHS of equation (\ref{kappa1}) vanish in the
large system limit and that only the auto-correlation part
(\ref{autocorrel}) contributes to the heat conductivity.

Before turning to numerical results in the next section, we should note
that our arguments though compelling do not constitute a formal proof that
the distributions 
(\ref{factorPtwoagain}) with $q$ symmetric are the unique stationary
solutions of equations (\ref{1Pevol}), (\ref{twopevol}) consistent with the
non-equilibrium boundary conditions. This is in fact a much deeper problem
that goes beyond the scope of the present paper. What we have shown
rather is that,
given that the system is in contact with thermal baths at distinct
temperatures, there exist stationary states of the form
(\ref{factorPtwoagain}) with associated heat current obeying
Fourier's law (\ref{currentdeltaT}) with the corresponding conductivity
(\ref{kappaness}). It is a natural conjecture to assume 
that, given the baths temperatures, the stationary state
(\ref{factorPtwoagain}) is unique. We turn to numerical simulations to
assess that claim.

\section{\label{sec.nr}Numerical computation}

Equations (\ref{kappanub}) and (\ref{kappaness}) can be verified
by direct numerical computations of the master equation
(\ref{mastereq}). This is achieved by adapting Gillespie's algorithm
\cite{Gil76,Gil77}, originally designed to simulate chemical or biochemical
systems of reactions, to our stochastic equation. The Monte-Carlo step here
necessitates two random trials. The first random number determines the time
that will elapse until the next energy exchange event. The second one
determines which one out of all the possible pairs of cells will perform an
exchange of energy and how much energy will be exchanged between them.

Thus, given a configuration $\{\epsilon_1,\dots,\epsilon_N\}$ of energies
at each one of the $N$ cells of the system, and, whether in the presence of
periodic boundary conditions 
(PBC), used to simulate an isolated system, or thermal boundary conditions
(TBC), used to simulate a system with both ends in contact with thermostats
at distinct temperatures $T_-$ and $T_+$, and, in that case, given
a pre-specified bath relaxation rate $\nu_\mathrm{Bath}$ associated with the
thermal baths, the time to the next collision is a random number with
Poisson distribution whose relaxation rate is specified by 
the sum of all collision frequencies,
\begin{equation}
  \eqalign{
    \gamma = \sum_{n=1}^{N-1} \nu(\epsilon_n,\epsilon_{n+1}) 
    + \nu(\epsilon_N,\epsilon_1)\,, \quad(\mathrm{PBC})\,,\\
    \gamma = \nu_\mathrm{Bath} \left(\sqrt{T_-} + \sqrt{T_+}\right)
    + \sum_{n=0}^{N} \nu(\epsilon_n,\epsilon_{n+1})\,,
    \quad(\mathrm{TBC})\,.
  }
  \label{rate}
\end{equation}
Given the rate $\gamma$, one draws a second random number $\chi$ uniformly
distributed on $(0,\gamma)$, and determines the pair $(n,n+1)$ of cells
which interact according to 
\begin{equation}
  \eqalign{
    \sum_{i=1}^{n-1} \nu(\epsilon_i,\epsilon_{i+1}) < \chi
    <   \sum_{i=1}^{n} \nu(\epsilon_i,\epsilon_{i+1})\,,
    \quad(\mathrm{PBC})\,,\\
    \lo\nu_\mathrm{Bath} \sqrt{T_-} + \sum_{i=0}^{n-1}
    \nu(\epsilon_i,\epsilon_{i+1}) < \chi  < \nu_\mathrm{Bath} \sqrt{T_-}  
    + \sum_{i=0}^{n} \nu(\epsilon_i,\epsilon_{i+1})\,, 
    \quad(\mathrm{TBC})\,.
  }
  \label{pair}
\end{equation}
In the latter equation, one understands that $\chi < \nu_\mathrm{Bath}
\sqrt{T_-}$ means updating the energy of the left bath, also associated with
$i=0$, and, similarly, $\chi> \nu_\mathrm{Bath} \sqrt{T_-}  
    + \sum_{i=0}^{N} \nu(\epsilon_i,\epsilon_{i+1})$  means updating the
    energy of the right bath, associated with cell $N+1$. How much energy is
    exchanged between the cells $n$ and $n+1$ is then 
determined by finding the energy $\eta$ which solves the equation 
\begin{equation}
  \int_{-\epsilon_{n}}^\eta  \ud \eta' W(\epsilon_{n}, \epsilon_{n+1}| 
  \epsilon_n - \eta', \epsilon_{n+1} + \eta') = 
  \chi - \sum_{i=1}^{n-1} \nu(\epsilon_i,\epsilon_{i+1})\,.
  \label{etasolve}
\end{equation}
Here we assumed PBC. The transposition to TBC is immediate.
For the sake of solving this equation, going back to the expression of the 
collision frequency
(\ref{collfrequencydef})-(\ref{collfrequency}), we compute the partial
integrals 
\begin{eqnarray}
  \fl
  \frac{1}{\nu(\epsilon_a,\epsilon_b)} 
  \int_{-\epsilon_b}^\eta  \ud \eta' W(\epsilon_a, \epsilon_b| \epsilon_a -
  \eta', \epsilon_b + \eta')\nonumber
  \\
  \lo= \left\{
    \begin{array}{l@{\quad}l}
      \sqrt{\frac{\epsilon_a}{\epsilon_b}}
      \frac{E\left(\frac{\eta+\epsilon_b}{\epsilon_a}\right) -
         \frac{\epsilon_a - \eta - \epsilon_b}{\epsilon_a}
         K\left(\frac{\eta+\epsilon_b}{\epsilon_a}\right)}
       { 2 E\left(\frac{\epsilon_a}{\epsilon_b}\right)
         - \frac{\epsilon_b - \epsilon_a}{\epsilon_b}
         K\left(\frac{\epsilon_a}{\epsilon_b}\right)}\,,&
      -\epsilon_b < \eta < \epsilon_a - \epsilon_b\, ,\\
      \sqrt{1+\frac{\eta} {\epsilon_b}}
      \frac{E\left(\frac{\epsilon_a}{\epsilon_b + \eta}\right)}
      { 2 E\left(\frac{\epsilon_a}{\epsilon_b}\right)
        - \frac{\epsilon_b - \epsilon_a}{\epsilon_b}
        K\left(\frac{\epsilon_a}{\epsilon_b}\right)}\,,&
      \epsilon_a - \epsilon_b < \eta < 0\, ,\\
      1 - \frac{
        E\left(\frac{\epsilon_a-\eta}{\epsilon_b}\right)
      - \frac{\epsilon_b - \epsilon_a + \eta}{\epsilon_b}
        K\left(\frac{\epsilon_a-\eta}{\epsilon_b}\right)}
      { 2 E\left(\frac{\epsilon_a}{\epsilon_b}\right)
        - \frac{\epsilon_b - \epsilon_a}{\epsilon_b}
        K\left(\frac{\epsilon_a}{\epsilon_b}\right)}\,,&        
      0<\eta<\epsilon_a\,,
    \end{array}
  \right.
  \label{partint}
\end{eqnarray}
and use a root finder routine --in our case the subroutine \texttt{rtflsp}
in \cite{NR}-- in order to
find the solution $\eta$ to equation (\ref{etasolve}).

We notice that the elliptic $K(x)$ functions which appear in equation
(\ref{partint}), are multiplied by $1-x$, which takes care of the
logarithmic divergence of $K$ at $x=1$, 
\begin{equation}
\lim_{x\to1} \left[K(x) - \frac{1}{2}\log\frac{16}{1-x}\right] = 0\,.
\end{equation}

\subsection{Periodic Boundary Conditions}

\begin{figure}[htpb]
  \centerline{
    \includegraphics[width=.8\textwidth]{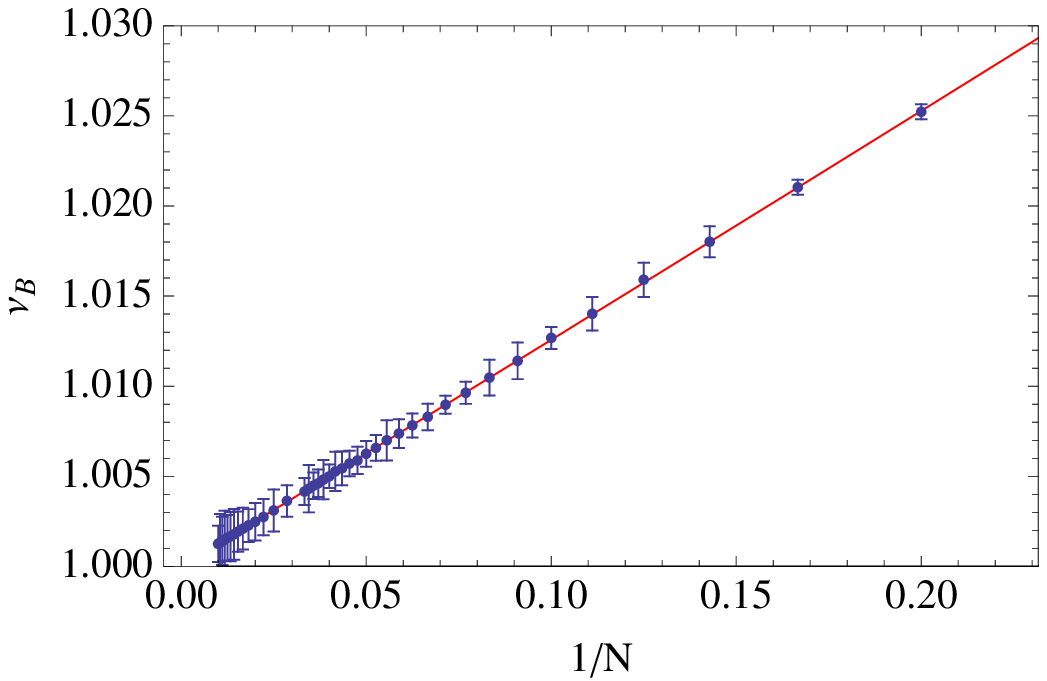}
  }
  \centerline{
    \includegraphics[width=.8\textwidth]{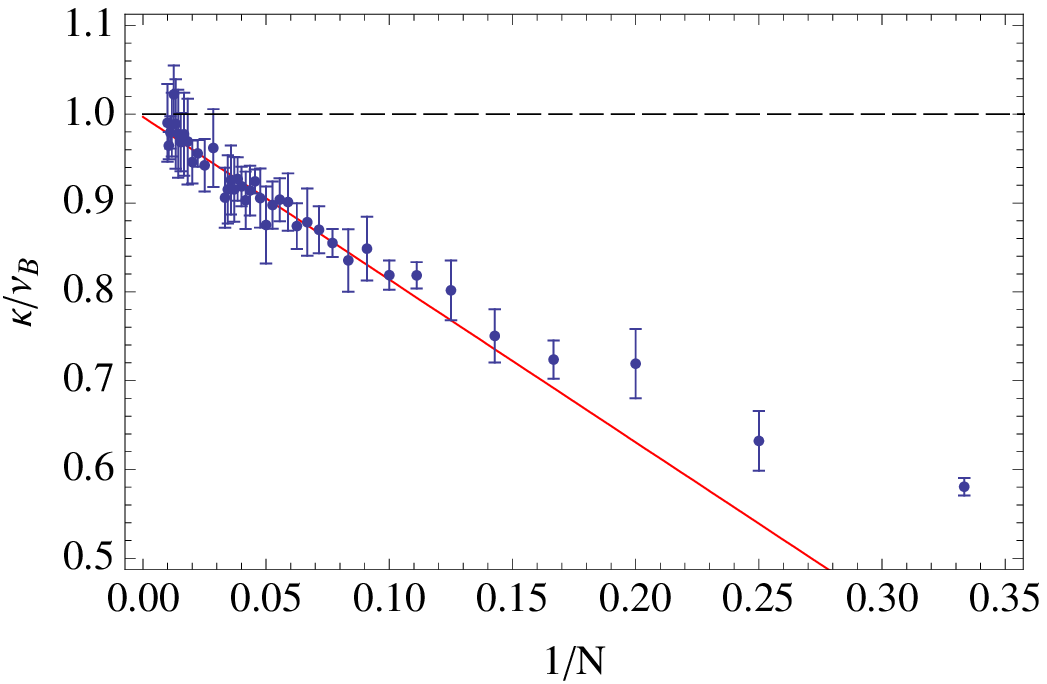}
  }
  \caption{(Top) Collision frequency computed under equilibrium conditions with
    energy $E = N$ and for different system sizes $N$, ranging from $N=3$
    to $N=100$, compared to the results of a numerical 
    integration of (\ref{nubN}) (solid red line). The $N\to\infty$
    asymptotic value is 1 according to (\ref{nub}). The agreement is
    spectacular. (Bottom) RHS of equation (\ref{kappahelf}), computed for
    the same parameters set as above. The ratio $\kappa/\nub$ is the
    infinite $N$ limit of these data, 
    which can be evaluated by linear regression (solid red line). Fitting
    all the data points corresponding to $N\ge 10$ and weighting the data
    according to the sizes of their error bars, we obtain $\kappa/\nub =
    0.997 \pm 0.004$.}  
  \label{fig.pbc}
\end{figure}

As a simple test of the algorithm, we compute the equilibrium average of the
collision frequency and compare it to the results of a numerical
integration of equation (\ref{nubN}) for different system sizes and energy per
particle taken to be unity. The results, displayed in the top panel of
figure \ref{fig.pbc}, are in excellent agreement.

Consider then the Helfand moment (\ref{defhelf}). We present in the bottom 
panel of figure \ref{fig.pbc} the results of numerical  
computations of equation (\ref{kappahelf}), obtained for different sizes $N$
of the system, using the numerical scheme described above. As can be seen
from the figure, $\lim_{n\to\infty} 1/(2N)  
\langle \Delta H(\tau_n)^2/\tau_n \rangle,$ has $\mathcal{O}(1/N)$
corrections to $\nub$ which are due to cross-correlations of the form 
$\langle \eta(\epsilon_{k_i}, \epsilon_{k_i+1}) \eta(\epsilon_{k_j},
\epsilon_{k_j+1}) \rangle$ that persist for finite $N$. The infinite $N$
extrapolation of our data is obtained by linear regression and yields
\begin{equation}
  \kappa/\nub = 0.997 \pm 0.004\,,
  \label{kappanuPBC}
\end{equation}
in precise agreement with equation (\ref{kappanub}) and our argument that
only static correlations contribute to the thermal conductivity.

\subsection{Thermal Boundary Conditions}

\begin{figure}[htpb]
  \centerline{
    \includegraphics[width=.5\textwidth]{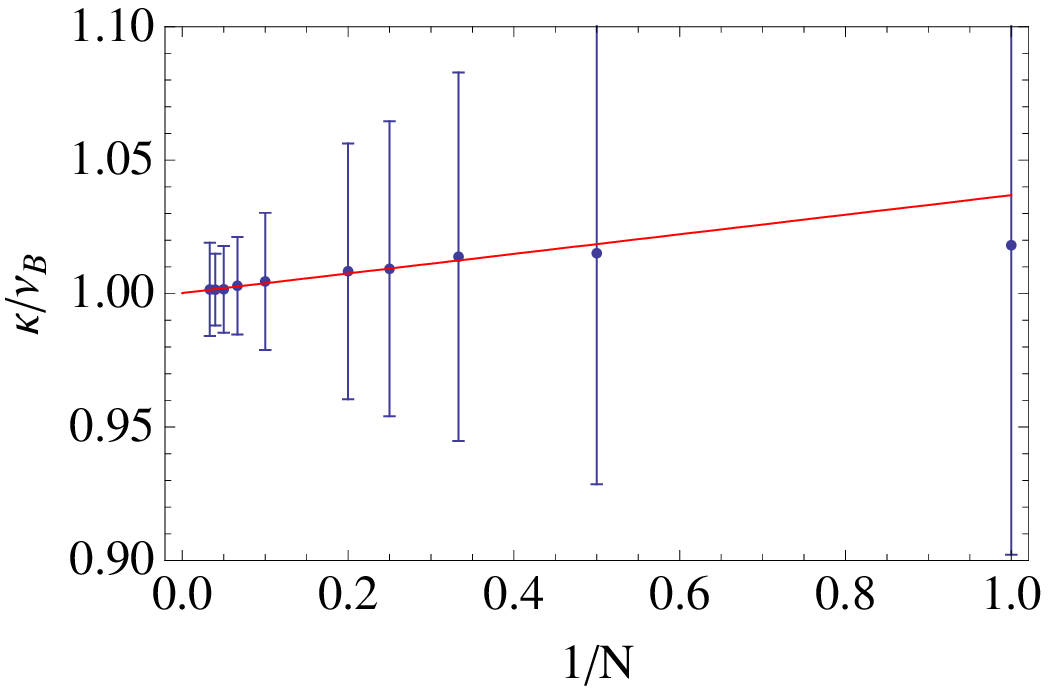}
    \includegraphics[width=.5\textwidth]{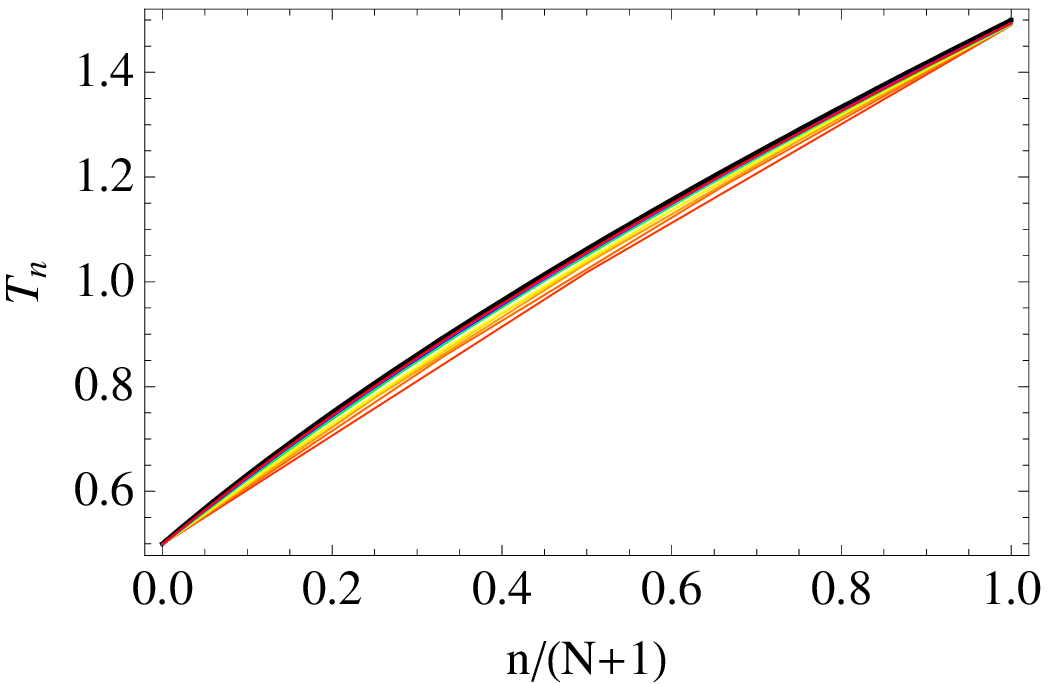}
  }
  \caption{(Left) Ratio between the heat current (\ref{currenttwop}) and
    local temperature gradient under thermal boundary conditions at 
    temperatures $T_+ = 1.5$ and $T_- = 
    0.5$, computed for different system sizes $N$, ranging from $N=1$ to
    $N=30$. The ratio $\kappa/\nub$ is computed from the infinite $N$
    limit of these data, evaluated by linear regression (solid red
    line). Fitting the data points weighed according to the sizes of their
    error bars, we obtain $\kappa/\nub = 1.0002 \pm  3 \, 10^{-4}$. (Right)
    Corresponding temperature profiles with stochastically thermalised
    cells at $n=0$ ($T_-$) and $n = N+1$ ($T_+$). The thick black line
    corresponds to  profile (\ref{noneqTn}) 
    expected from Fourier's law. The data is barely distinguishable from
    this curve when $N$ is sufficiently large.}
  \label{fig.tbc}
\end{figure}

We simulate thermal boundary conditions according to the scheme described
above, with bath relaxation frequency $\nu_\mathrm{Bath} = 100$, fixing the
baths temperatures to $T_- = 0.5$ and $T_+ = 1.5$. The heat conductivity is
then evaluated according to equation (\ref{fourierlaw}) by computing the
average heat flux, divided by the temperature gradient, which is decreased
by increasing the system size $N$. The results, displayed on the left panel
of figure \ref{fig.tbc}, yield the ratio
\begin{equation}
  \kappa/\nub = 1.0002 \pm 3 \, 10^{-4}\,,
  \label{kappanuTBC}
\end{equation}
which provides a very nice confirmation of equation (\ref{kappaness}).

The corresponding temperature profiles which, according to Fourier's law,
are expected to be
\begin{equation}
T_n = \left[\frac{1}{2}(T_-^{3/2} + T_+^{3/2}) + \frac{n}{N+1}
(T_+^{3/2} - T_-^{3/2})\right]^{2/3}\,,
\label{noneqTn}
\end{equation}
are shown on the right panel of figure \ref{fig.tbc}.

\section{\label{sec.con}Conclusions}

To summarize, we have obtained a systematic derivation of Fourier's law and
computed the heat conductivity of a class of stochastic systems describing,
at a mesoscopic level, the rare energy exchanges in Hamiltonian systems
of two-dimensional confined particles in interaction that were introduced
in \cite{GG08a, GG08b}.

A remarkable feature of our approach is that the structure of the
stationary state is determined by the geometry of the system and not
by the actual form of the kernel which specifies the nature of the
interactions between the system cells. Indeed our derivation of the
two-cell distribution function only used the nearest neighbour nature of
the interaction. This property alone justifies that the leading part of the
distribution function, meaning including linear order in the local
temperature gradients, is the local thermal equilibrium plus a two-cell
function, symmetric with respect to its arguments. What precise form this
function has depends on the actual process under consideration. Whatever
this form, its symmetry suffices to justify that the only contributions to
the heat current come from the local thermal equilibrium part, at least to
the extent that the current is computed to linear order in the local
temperature gradients. 

We therefore conclude the same property, namely that the heat conductivity
can be obtained through averages with respect to the local thermal
equilibrium distributions, should hold for systems similarly described by
an equation of the form (\ref{mastereq}), sustaining non-equilibrium
stationary states with symmetric corrections of the form
(\ref{factorPtwoagain}).

In particular, we observe that, whether or not such systems obey the
gradient condition, the transport coefficient is given from the diffusion of 
Helfand moment, or equivalently the Green-Kubo formula, through a static
average only. 

Let us mention that, as observed  in \cite{GL08}, the class of mechanical
systems with confined particles in interaction that share the transport
properties of the systems considered in \cite{GG08a, GG08b} is larger than
the class of semi-dispersing billiards envisaged there. In particular, the
systems studied in \cite{GL08} do not \emph{a priori} obey a master
equation of the form (\ref{mastereq}). We expect that the BBGKY hierarchy
applied to the phase-space distributions of these systems will have
solutions with properties similar to the solutions found here, namely a
local thermal equilibrium part plus a symmetric correction which does not
contribute to the heat current.

\ack
The authors are indebted to J.~R. Dorfman for sharing his expertise on the
Chapman-Enskog expansion with them. They also acknowledge useful
discussions with J. Bricmont, M. D. Jara Valenzuela, R. Lefevere,
C. Liverani and S. Olla.  This research is financially supported by the
Belgian Federal Government under the Interuniversity Attraction Pole
project NOSY P06/02 and the Communaut\'e fran\c{c}aise de Belgique under
contract ARC 04/09-312. TG is financially supported by the Fonds de la
Recherche Scientifique F.R.S.-FNRS.

\end{document}